\documentclass[12pt]{article}
\usepackage{amsmath,amssymb}

\usepackage{graphicx}
\usepackage{wrapfig}
\usepackage{color}

\usepackage{hyperref}
\usepackage[sort,compress]{cite}

\def\beq{\begin{eqnarray}}
\def\eeq{\end{eqnarray}}

\def\lb{\label}

\newcommand{\Tr}{\,\mathrm{Tr}\,}            

\newcommand{\be}{\begin{equation}}
\newcommand{\ee}{\end{equation}}
\newcommand{\bea}{\begin{eqnarray}}
\newcommand{\eea}{\end{eqnarray}}
\newcommand{\bg}{\begin{gather}}

\newcommand{\bseq}{\begin{subequations}}
\newcommand{\eseq}{\end{subequations}}

\renewcommand{\ln}{\mathop{\rm ln}\nolimits}

\def\tr{\hbox{Tr}}

\def\be{\begin{eqnarray}}
\def\ee{\end{eqnarray}}
\def\lb{\label}

\begin{document}

\title{\textbf{
\rm Manifestations of $(2+1)d$ chiral 
anomaly \\
in a graphene plate
}}
\vspace{1cm}
\author{ \textbf{  Sergey N. Solodukhin}} 

\date{}
\maketitle
\begin{center}
\emph{Institut Denis Poisson, \\
UMR-CNRS 7013, Universit\'e de Tours, \\
Parc de Grandmont, 37200 Tours, France}


\end{center}




\vspace{0.2mm}

\begin{abstract}

\noindent { 

Inspired by  the Dirac model model of graphene,  we consider a  $(2+1)$-dimensional   fermionic system in which fermions are described by four-component spinors. These fermions are proposed to interact with an electromagnetic field originating from a four-dimensional setting, as the graphene plate is embedded in 4d Minkowski spacetime. In this framework, a chiral anomaly arises at the boundary of the plate, stemming from a non-local anomaly action that depends on both the electromagnetic and chiral gauge fields when the chiral transformation is localized. This results in boundary chiral and electric currents, and we explore potentially observable effects when external magnetic or electric fields are applied to the fermionic system.}

\end{abstract}

\vskip 1 cm
\noindent
\rule{7.7 cm}{.5 pt}\\
\noindent 
\noindent

\noindent ~~~ {\footnotesize e-mail: sergey.solodukhin@univ-tours.fr}

\pagebreak

\noindent{\it Introduction.}  
Quantum anomalies are becoming increasingly significant in physical systems that are accessible to direct experimental studies. One experimental setting is the quark-gluon plasma. Several such systems are also known in condensed matter, such as Dirac semimetals and graphene. The latter is particularly interesting as it is confined to a plane and can be effectively described by $(2+1)$-dimensional field theory models. For a review of recent developments in these areas, see \cite{Chernodub:2021nff}.
In those systems,  the chiral anomaly contributes to   conserved quantities such as electric current \cite{Alekseev:1998ds}, \cite{Fukushima:2008xe} and 
the stress energy tensor \cite{Solodukhin:2024gjx}, which may have observable consequences.

The main goal of this note is to discuss the possible experimental manifestations of the chiral anomaly in $(2+1)$-dimensional fermionic systems. This anomaly has been recently studied theoretically in \cite{FarajiAstaneh:2023fad}. One peculiarity of this anomaly is that it appears at the boundary of the $(2+1)$-dimensional system. Another notable feature is that four-component fermions are required to properly formulate the boundary conditions and to define well-behaved chiral transformations. In the two-component fermion representation, chirality is not well defined
since a chirality matrix  does not exists. This motivates the use of the four-component representation for the fermions where the chirality matrix can be defined.
For the sake of completeness,  below we will briefly review the formulation of the boundary conditions and the computation of the anomaly, closely following  \cite{FarajiAstaneh:2023fad}.

It should be noted that a similar, though conceptually different, parity anomaly has been previously discussed in the literature (see \cite{Niemi:1983rq}). 
In this note, we do not address the parity anomaly. 
There have also been earlier works \cite{CA} discussing the appearance of an electric current in four dimensions due to gauge field terms in the conformal anomaly.
In the three-dimensional case considered here, there is no conformal anomaly in the bulk, though an anomaly may appear on the boundary, as shown in \cite{FarajiAstaneh:2023fad}. However, the background gauge field does not contribute to the boundary anomaly, and thus, in three dimensions, there is no effect analogous to that discussed in \cite{CA}.
The origin of the emergent currents in our case is the boundary chiral anomaly, as we now demonstrate.

\bigskip

\noindent{\it The model.}  In this note we consider the $(2+1)$-dimensional spacetime $M_3$ as embedded in a bigger $(3+1)$-dimensional Minkowski spacetime $M_4$.
We use the units in which speed of light  $c=1$.
Using the Cartesian coordinates $(X_0=t, X_1, X_2, X_3)$ in $M_4$  we define $M_3$ as  hypersurface where $X_3=0$. Additionally, spacetime $M_3$ is supposed to have a boundary
$\partial M_3$ that can be defined by condition $x^n=n^i x_i=const$, where $(x_i\, , \, i=0,1,2)$ are the intrinsic coordinates in  $M_3$ and $n^i, i=0,1,2$ is the normal vector to the boundary
in $M_3$. 

As it was announced above we will consider the four-component fermions $\psi^{\bf a}$ with a colour index ${\bf a}=1, \dots, N$.  For graphene $N=2$. The fermions are  coupled to the Maxwell field $A_i, i=0,1,2$ 
that has the four-dimensional origin, i.e. it is a part of the four-dimensional Maxwell field $A_\mu, \mu=0,1,2,3$ defined in Minkowski spacetime $M_4$. The Maxwell field $A_\mu$ is considered to be a classical background field for the quantum fermions. For a similar setup see \cite{Fialkovsky:2012ee}.
The model includes the massless fermions.  In its most general form,   the action for the massless  fermions in three dimensions reads
\be
W=\int_{M_3}dt d^2x \, h\,  \sum_{{\bf a}=1}^N\bar{\psi}^{\bf a}\hat{D}\psi^{\bf a} \, , \  \  \hat{D}= i\hat{\gamma}^k(D_k -i eA_k)\, ,
\lb{1}
\ee
where $\bar{\psi}=\psi^\dagger \gamma^0$, $\hat{\gamma}^k=\sum_{p=0}^2 h^k_p \gamma^p$ and index $k$ taking values $0, 1,2$. Here, $h^k_p$ are the $3$-beins and $h^{-1}=\det (h^k_p)$, which generally describe   the curved spacetime $M_3$, and $D_k$ is the respective covariant derivative defined in termes of the Lorentz connection.
The matrices  $\gamma^0, \gamma^1$, $\gamma^2$, along with matrix $\gamma^3$, form the standard   $4\times 4$ gamma-matrices $\{ \gamma^\mu\,  ,\mu=0,.., 3\}$ defined in $M_4$.  
For the graphene plate, we have $h^0_0=v_F^{-1}$ and $h^1_1=h^2_2=1$, so that $h=v_F$, where $v_F\approx c/300$ is  the Fermi velocity.

The use of Dirac fermions as in (\ref{1})  to describe graphene has a long story and  it is known as the Dirac model of graphene \cite{DM}.
We were in particular motivated by Son's paper \cite{Son:2007ja} with the modification that  we neglect the mutual quantum  Coulomb  interaction between fermions. Instead,  the fermions are coupled to external  classical Maxwell fields which are   four-dimensional and satisfy the four-dimensional Maxwell equations. One could also add the proper four-dimensional action   for $A_\mu$  to  (\ref{1}).
For a related quantum model see \cite{CH}.

The model (\ref{1}) retains  the usual four-dimensional chiral symmetry,  $\psi' = e^{-i\alpha \gamma_5}\psi$,  generated by $\gamma_5=-i \gamma^0\gamma^1\gamma^2\gamma^3$ matrix. 

\bigskip

\noindent{\it Boundary conditions.} One has to specify the boundary conditions to be imposed on the fermionic field at the boundary $\partial M_3$. It has been discussed in detail in
\cite{VF} and in the case at hand in \cite{FarajiAstaneh:2023fad}.  
We will briefly review the reasoning in \cite{FarajiAstaneh:2023fad} that leads to these conditions. In the context of graphene the boundary conditions were discussed earlier in \cite{Biswas:2022dkg} and \cite{Herzog:2018lqz}.
Near the boundary we split the bulk coordinates $\{ x^i\,, i=0,1,2\}=\{x^n, x^a\}$  on intrinsic coordinates $x^a, a=0,1$ and the coordinate $x^n$ normal to the boundary $\partial M_3$.

 A natural condition is to require that
the normal component of the fermionic current vanishes on the boundary, $\psi^\dagger\gamma^0\gamma^n\psi|_{\partial M_3}=0$, $\gamma^n=n_k \gamma^k$. This can be achieved by imposing the Dirichlet condition on half of the components, $\Pi_-\psi|_{\partial M_3}=0$, where we have introduced a pair of projectors $\Pi_+$ and $\Pi_-$, $\Pi_++\Pi_-=1$. The normal component of the current vanishes, provided that $\Pi_+\gamma^n=\gamma^n \Pi_-$. However, this condition alone is insufficient; we must   impose an additional condition on the other half of the components of $\psi$.
This necessity arises from the requirement that the imposed conditions should hold for the eigenvalues of the Dirac operator, $\hat{D}\psi=\lambda \psi$.
This can be illustrated in a simple case when the boundary $\partial M_3$ is a plane.
This leads to a new condition that $\Pi_- (i\hat{\gamma}^k\partial_k\psi)|_{\partial M_3}=0$.
Separating the Dirac operator on the normal and tangential directions, $i\hat{\gamma}^k\partial_k=i\hat{\gamma}^n\partial_n+i\hat{\gamma}^a\partial_a$, one finds, after some algebra, that one has to impose the Robin type  boundary condition $\partial_n \Pi_+\psi|_{\partial M_3}=0$. One also finds additional commutation relation between the projectors and  the gamma matrices,
$\Pi_-\gamma^a=\gamma^a \Pi_-$, in the tangential directions. One can represent $\Pi_\pm=\frac{1}{2}(1\pm \chi)$, where matrix $\chi$ has to commute with $\gamma^a$ and
anti-commute with$\gamma^n$. With the gamma-matrices at hand this can be achieved by choosing\footnote{A more general choice would be $\chi(\theta)=i\gamma^3\gamma^ne^{i\theta \gamma^3}$, $\chi^2(\theta)=1$,  similarly to 
the chiral bag boundary condition in four dimensions  \cite{GK}, \cite{Ivanov:2021yms}, \cite{Herzog:2018lqz}.   In four dimensions using the trick of Gilkey and Kirsten \cite{GK} one shows that the respective heat kernel coefficient and the
chiral anomaly do not depend on $\theta$. This trick goes through in the three-dimensional case at hand for standard 3d Dirac operator thus indicating that parameter $\theta$ does not show up in the chiral anomaly (switching on the chiral gauge field may require a more careful analysis). In the rest of the paper we use $\chi(0)$  in the boundary condition.}    
 $\chi=i\gamma^3 \gamma^n$. Matrix $\chi$ can be represented in a slightly different form
using the fact that $M_3$ is embedded in a larger space $M_4$. Indeed, from the four-dimensional point of view the boundary $\partial M_3$ is a co-dimension two hypersurface with two
normal vectors, one of which is vector $n$ and the other is vector $k$ with only one non-vanishing component $k_3$. In terms of these two vectors one may define a binormal  as $\epsilon_{\mu\nu}=\frac{1}{2}(k_\mu n_\nu-k_\nu n_\mu)$. The matrix $\chi$ then can be written as $\chi=i\epsilon_{\mu\nu}\gamma^\mu \gamma^\nu$.

In the presence of gauge fields and in the case of a non-planar boundary and/or curved  $(2+1)d$ spacetime, the entire procedure can be repeated and leads to a modified Robin boundary condition, 
$(\nabla_n-S)\Pi_+\psi|_{\partial M_3}=0$, where exact form of $S$ is given in Appendix. To summarise, one imposes the following boundary conditions  \cite{Vassilevich:2003xt}, \cite{VF},  \cite{FarajiAstaneh:2023fad},
\be
(\nabla_n-S)\Pi_+\psi\oplus \Pi_- \psi =0
\lb{2}
\ee
for the fermionic field on the boundary $\partial M_3$.   

A word of caution is in order. The discussion here is an idealization, while the realistic conditions in graphene depend on the specific placement of the boundary within the hexagonal lattice.


\bigskip

\noindent{\it Heat kernel coefficients and the chiral anomaly.}
It is convenient to make the chiral transformation local, $\delta \psi =-i\alpha(x)\gamma_5 \psi$. To preserve this invariance, a corresponding gauge field $A^5_k$ is introduced, which transforms as $\delta A^5_k=\partial_k \alpha(x)$. The respective Dirac operator  to be considered in action (\ref{1}) is then given by
\be
\hat{D}_5=i\hat{\gamma}^k(D_k-ieA_k +iA^5_k\gamma_5)\, .
\lb{3}
\ee
 The variation of the action with respect to the chiral gauge field defines the chiral current, $j^k_5=\frac{\delta W}{\delta A^5_k}$, which is classically conserved $\nabla_k j^k_5=0$, provided the fermions satisfy the Dirac equation, $\hat{D}_5\psi^{\bf a}=0$. After quantizing the fermions, the system is described by the quantum action, $W_Q=-\frac{1}{2}\ln \det \hat{D}_5^2$. Its variation under local chiral transformations then is given by equation \cite{Vassilevich:2003xt},
\be
\delta_\alpha W_Q=-2i\, a_3(\gamma_5\alpha(x), \hat{D}_5^2)\, .
\lb{4}
\ee
The non-invariance of the quantum action indicates the presence of a quantum chiral anomaly.
 Generally, in dimension $d$, for a matrix valued function $Q$ and a Laplace type differential operator $\hat{D}^2$ the coefficients $a_k$ are defined as $\Tr( Q e^{-s\hat{D}^2})=\sum_{k\geq 0}s^{(k-d)/2}a_k(Q,\hat{D}^2)$. These coefficients are locally computable.  In odd dimension $d$ the coefficient $a_d$ is due to the boundary terms only.
 We refer to \cite{Vassilevich:2003xt} for the general results for these coefficients and to \cite{FarajiAstaneh:2023fad} for a computation in the 3d case at hand. Note, however, that in  \cite{FarajiAstaneh:2023fad}  the chiral gauge field   was not included in the Dirac operator.  The details of the calculation for the
 Dirac operator in the form (\ref{3}) are provided in the Appendix. We now state the result,
 \be
 \delta_\alpha W_Q=\frac{eN}{4\pi}\int_{\partial M_3} \alpha (x) \epsilon^{ab}F_{ab}\, , \  \  F_{ab}=\partial_a A_b-\partial_b A_a
 \lb{5}
 \ee
 The chiral anomaly arises solely from the boundary terms, as was earlier obtained in  \cite{FarajiAstaneh:2023fad}. 
 Generelizing  \cite{FarajiAstaneh:2023fad}, we find that the anomaly is independent of the chiral gauge field $A^5_k$. 
 Here, the boundary epsilon symbol is defined as $\epsilon^{ab}=\epsilon^{nab3}$  (we use convention $\epsilon_{01}=-1$)
 in terms of the normal projection of the  four-dimensional symbol. It is worth noting that integral in (\ref{5}) can be written in terms of differential form $F=dA$ as
 $\int_{\partial M_3}\alpha(x) F$, and  is therefore  independent of the induced metric, or the $2$-beins, on $\partial M_3$.  Consequently, the chiral anomaly (\ref{5})  does not depend on the Fermi velocity $v_F$,
 which appears through the $3$-beins in the action (\ref{1}).
 
 \bigskip
 
 \noindent{\it Anomaly action and  local form of the anomaly.} 
Although the chiral gauge field \( A^5_k \) does not directly contribute to the chiral anomaly (Eq. \(\ref{5}\)), it can still be used to integrate the anomaly. Assuming the quantum action depends on both the gauge field \( A_k \) and the chiral gauge field \( A^5_k \), we can readily find a gauge invariant term whose variation reproduces the anomaly (\ref{5}):
\be
W_{\rm anom} = -\frac{eN}{4\pi} \int_{\partial M_3} A^5_a \epsilon^{ab} \left(A_b - \partial_b \Box^{-1}_{(2)} \nabla^c A_c \right),
\lb{6}
\ee 
where \( \Box_{(2)}=\nabla^a\nabla_a \) is the intrinsic wave operator along the two-dimensional boundary $\partial M_3$. We  emphasize that this is only the part of the quantum action  responsible for the chiral anomaly. The total action also includes terms that do not depend on $A^5_k$.
 The anomalous chiral current is obtained by varying the anomaly action with respect to $A^5_k$. 
  In the bulk, this current is conserved, but at the boundary, its intrinsic divergence is non-vanishing,
 \be
 &&\nabla^k j^5_k(x)=0\  \  \  {\rm if}  \  x\ \  {\rm lies } \  \  {\rm inside} \  M_3\, , \nonumber \\
 &&\nabla^a j^5_a(x)=-\frac{eN}{8\pi}\epsilon^{ab}F_{ab}\, \delta_{\partial M_3} 
 \lb{7}
 \ee
 The anomaly thus manifests  \cite{FarajiAstaneh:2023fad}  as the non-conservation of the  chiral current at the boundary.

 It is worth noting that the boundary anomaly (\ref{7}) has a form similar to the chiral anomaly in two spacetime dimensions, discussed, for instance, 
 in \cite{Alekseev:1998ds}. A significant difference, however, is that in the two-dimensional case, the fermions have only two components, which establishes an algebraic relation between 
 the electric and chiral currents. In the present case, with four-component fermions, such a relation does not exist. Another important distinction is that in two dimensions, 
 the chiral gauge field contributes to the chiral anomaly alongside with the Maxwell gauge field, as demonstrated in \cite{Marachevsky:2003zb}.

\bigskip

\noindent{\it Boundary chiral and electric currents.}  The tangential component of the chiral current defines the boundary current  as
$j^5_a=J^5_a\delta_{\partial M_3}$.  Similarly,  the tangential component of electric current can be obtained by varying with respect to the electric field $A_a$,
yielding  $j_a=J_a\delta_{\partial M_3}$. From the action (\ref{6}) one can then derive the expressions for the boundary currents,
\be
&&J^5_a=-\frac{eN}{4\pi}\epsilon_{ab}( A^b-\partial^b \Box^{-1}_{(2)}\nabla^c A_c)\, , \  \  \nabla^a J^5_a=-\frac{eN}{8\pi}\epsilon^{ab}F_{ab} \nonumber \\
&&J_a=\frac{eN}{4\pi}(\epsilon_{ab} A_5^b-\partial_a \Box^{-1}_{(2)} {\cal F})\, , \  \  \  \  \  \nabla_ a J^a=0
\lb{8}
\ee
where we define ${\cal F}=\epsilon^{ab}\partial_a A^5_b$.
Notice that both electric current $J_a$ and the chiral current $J^5_a$ are gauge invariant due to the presence of  non-local terms.

It is typically assumed, and we will adopt this here,  that only the temporal component $A^0_5=\mu_5$ of the chiral gauge field is non-vanishing and, furthermore, constant.
Its value is interpreted as the chiral chemical potential $\mu_5=\frac{1}{2}(\mu_L-\mu_R)$. 
More generally, one may impose the constraint ${\cal F}\equiv\epsilon^{ab}\partial_a A^5_b=0$,
which allows the expression for the electric current to be made local. Additionally, this constraint localizes the anomaly action (\ref{6}), 
\be
W_{\rm anom} = -\frac{eN}{4\pi} \int_{\partial M_3} A^5_a \epsilon^{ab} A_b\, .
\lb{8-2}
\ee

With these constraints, it follows that the temporal component of the boundary  electric current vanishes, $J_0=0$, leaving only the spatial component non-vanishing. This defines 
the spatial vector $\vec{J}=J_1\vec{e}_1$, where vector $\vec{e}_1$ is tangent to the boundary. The component $J_1$  remains  constant along the boundary, 
\be
J_1=\sigma \mu_5\, , \  \  \  \sigma=\frac{eN}{4\pi}\, .
\lb{9}
\ee
On the other hand, one finds for the components of the chiral current,
\be
J^5_0=\sigma (A_1-\partial_1 \chi)\, ,  \  \   \   J^5_1=\sigma (A_0-\partial_0\chi)\, , \  \  \   
\lb{9-1}
\ee
where the scalar field $\chi$ is a solution to the equation $ \Box_{(2)}\chi=\nabla^c A_c$.

\bigskip

In the remainder of this note, we will discuss several possible manifestations of the $(2+1)d$ chiral anomaly. The  $(2+1)$-dimensional fermionic system  is identified with 
a graphene plate that has a non-trivial boundary. 

\bigskip

\noindent{\it Graphene disk  in a constant magnetic field.}  In this scenario, we consider a graphene disk, denoted by \(\cal D\), placed in an external magnetic field \(\vec{B}\). The disk is taken to be round for simplicity, and it is situated in the \(z=0\) plane of a cylindrical coordinate system \((z, r, \phi)\), where the radial coordinate \(r\) is constrained to \(r=R\), defining the radius of the disk. The boundary of the disk, \(\cal C\), is a closed curve, which in this case is a circle with radius \(R\).

The interesting effect here is that the external magnetic field \(\vec{B}\) induces a chiral charge on the boundary \(\cal C\) of the graphene disk. To demonstrate this, we use the relation 
$J^5_0=\sigma (A_1-\partial_1\chi)$, where \(J^5_0\) represents the chiral charge density on the boundary  and \(A_1\) is the component of the background gauge field parallel to the boundary. Here, we define \(x^1=R\phi\) as the coordinate along the boundary, which corresponds to the azimuthal direction around the disk.

By recognizing that the Maxwell gauge field originates from a four-dimensional context, we can apply the four-dimensional Maxwell equations to our system. 
The total chiral charge induced on the boundary \(\cal C\) of the graphene disk is given by the integral:
\be
Q_5({\cal C})=\oint_{\cal C}J_0^5=\sigma \oint_{\cal C} \vec{A} d\vec{x} =\sigma\int_{\cal D} \textup{curl} \, \vec{A} \,  d\vec{S}
\lb{10}
\ee
Here, the gauge field \(\vec{A}\) is integrated along the boundary \(\cal C\), and through Stokes' theorem, this can be converted into a surface integral over the disk \(\cal D\), involving the curl of \(\vec{A}\), which gives the magnetic field \(\vec{B} = \textup{curl} \, \vec{A}\). Thus, the total chiral charge is related to the magnetic flux \(\Phi(\cal D)\) passing through the disk:
\be
Q_5({\cal C})=\sigma\Phi({\cal D})\, , \quad  \Phi({\cal D})= \int_{\cal D} \vec{B} \cdot d\vec{S}
\lb{11}
\ee
For a constant magnetic field \(\vec{B}\) that is perpendicular to the surface of the disk \(\cal D\), the magnetic flux \(\Phi(\cal D)\) becomes:
$
\Phi({\cal D}) = \pi  R^2 B\, ,
$
where \(R\) is the radius of the disk and \(B\) is the magnitude of the constant magnetic field. This result shows that the induced chiral charge on the boundary of the graphene disk is proportional to the magnetic flux passing through the disk. Consequently, a time-varying magnetic field results in a time-dependent variation of the induced chiral charge on the boundary of the plate. This variation can be expressed in terms of the circulation of the electric field as:
\be
\frac{dQ_5}{dt}=-\sigma \oint_{\cal C} \vec{E} d\vec{x}\, .
\lb{11-1}
\ee

\bigskip

\noindent{\it Graphene ring in magnetic field of a cylindrical solenoid.} 
In this setup, we consider a graphene ring placed in the magnetic field generated by a solenoid. The solenoid is positioned at the center of the ring along its axis of symmetry, meaning the magnetic field is directed along the axis of the ring in the cylindrical coordinate system $(z,r,\phi)$.   The graphene ring has two boundaries: an inner boundary  ${\cal C}_1$  and an outer boundary  ${\cal C}_2$ , both of which are concentric circles in the plane of the ring. The magnetic field inside the solenoid is uniform and parallel to the $z$-axis, and the vector potential $\vec{A}$ outside the solenoid decreases with distance,
\be
A_\phi=\frac{B a^2}{2 r}\, , \  \   r\geq a
\lb{13}
\ee
where $a$ is the radius of the solenoid.
Since there is no magnetic field outside the solenoid, the magnetic flux through the ring vanishes, which means the total chiral charge induced on both boundaries is zero. However, the vector potential  (\ref{13})  produces a circulation around each component of the boundary, leading to the respective chiral charge,
\be
Q_5[{\cal C}_2] =-Q_5[{\cal C}_1]=\sigma\pi B a^2\, .
\lb{14}
\ee 
Notice that the induced boundary chiral charge depends only on the intrinsic parameters $B$ and $a$  of the solenoid. 
Chiral charge $Q_5$ can be represented as $Q_5=n_L-n_R$, where $n_{L(R)}$ is quantum average of number of left- and right-handed fermions.
Physically, the discussed effect thus can be interpreted as a separation of chirality: fermions of predominantly one chirality concentrate on one component of the boundary, while fermions of the opposite chirality reside on the other component.

It is also interesting to note that this effect, much like the Aharonov-Bohm effect, arises from the manifestation of the gauge potential in regions where the magnetic field is absent.

\bigskip

 \noindent{\it Intrinsic magnetic field and magnetic moment of a graphene disk.}  As one of the manifestations of the chiral anomaly, we observe a permanent electric current along the boundary with a constant value, as shown in (\ref{9}). This current is independent of the external electric field and persists even in its absence, forming a circular current loop that produces magnetic field. 
 This is a text-book situation discussed, for instance, in \cite{Jackson}.  We can choose the spherical coordinates $(r, \theta, \phi)$  centered on the disk, ensuring the disk lies at $\theta=\pi/2$.  At the observation point far from the disk, the magnetic field produced by the loop is given by \cite{Jackson}
 \be
 B_r=2\pi R^2\sigma\mu_5 \frac{\cos\theta}{r^3}\, , \  \  \ 
 B_\theta=\pi R^2\sigma \mu_5 \frac{\sin\theta}{r^4}\, .
 \lb{15}
 \ee
 We can associate a magnetic dipole moment with the disk, 
 \be
 \vec{m}=\sigma \mu_5 \pi R^2\, \vec{k}\,  ,
 \lb{16}
 \ee
 as the product of the loop current and the disk area, $\vec{k}$ is vector normal   to the disk surface.  The magnetic moment of the graphene disk couples to an external magnetic field in the usual way.
This interaction could serve as yet another manifestation of the chiral anomaly, accessible for direct observation.
 
The following remark is in order.  In the above consideration the boundary electric current  (\ref{9}) with constant intensity proportional to $\mu_5$
 is considered to be non-dissipative as in a superconducting phase.  
 In real physical systems, however, the dissipation of energy perhaps should not be neglected.
 A further logical possibility  to be  worth mentioning here is that  in  isolated graphene samples, in order to avoid dissipation of energy  if there are no external fields, the chemical potential should vanish.
This issue requires a more careful analysis of equilibrium and non-equilibrium situations in graphene.

\bigskip

\noindent{\it Graphene disk/ring in electric field of linear charge.} 
Consider a static configuration of a linear electric charge distribution placed along the geometric axis of symmetry of the disk. 
In this configuration, the electric field \( \vec{E} = \frac{\lambda}{2\pi r} \vec{e}_r \) arises from a static linear charge distribution, where \( \lambda \) is the linear charge density, and the field is directed radially. The corresponding electric potential \( A_0(r) \) is derived from the relationship \( \vec{E} = -\vec{\nabla} A_0 \), yielding
\be
A_0(r) = - \frac{\lambda}{2\pi} \ln r,
\lb{16-1}
\ee
where \( r \) is the radial distance from the axis. Since the system is static and rotationally symmetric, the condition \( \partial^a A_a = 0 \) holds, implying a constant scalar function \( \chi \).

The spatial component of the boundary chiral current $\vec{J}_5$ is then 
$J^5_1 = \sigma A_0$, which remains constant along the circular boundary of radius \( R \). The total current is given by 
\be
I_5 =\oint_{\cal C} \vec{J}_5 d\vec{x} = -\sigma \lambda R \ln R + \text{const}\, .
\lb{17}
\ee
This indicates the appearance of a permanent chiral current at the boundary, whose intensity grows as \( R \ln R \) when the radius of the boundary increases.
Notice that in this configuration the boundary chiral current $\vec{J}_5$ is perpendicular to the electric field $\vec{E}$.
In a ring configuration, the current along the inner boundary of the ring flows in the direction opposite to that along the outer boundary.

\bigskip

\noindent{\it Conclusion.}  
In this note we consider a fermionic system in $(2+1)d$ spacetime with a boundary, where fermions are described by four-component spinors. This system provides an intriguing example of a boundary-induced chiral anomaly. When applied to a finite-sized graphene plate, our analysis suggests several experimental signatures of the anomaly, including an induced chiral charge and electric current localized on the boundary. These effects are expected to arise when the system is placed in the external magnetic or electric fields, leading to certain experimental predictions that could be tested.

\appendix
\section{Appendix}
\setcounter{equation}0
\numberwithin{equation}{section}
Here we consider the general case of a curved spacetime $M_3$ with a non-planar boundary $\partial M_3$.
For the general formulas we refer to review  \cite{Vassilevich:2003xt}, which we follow here.
The square of the 3d Dirac operator (\ref{3})
takes the form,
\be
\hat{D}_5^2=-(\nabla^i\nabla_i +E)\, ,
\lb{a1}
\ee
where  $\nabla_i=D_i^L+\omega_i$, $D_i^L$ is the usual Lorentz connection and 
\be
\omega_i=ie A_i+\frac{i}{2}[\gamma_i,\gamma_j] \gamma_5 A_5^j
\lb{aa}
\ee
is an additional term, see \cite{Vassilevich:2003xt}. The difference in two covariant derivatives leads to an extra term in
the Robin boundary condition, 
\be
(\nabla_n-S)\Pi_+\psi=0\, , \  \ S=(-\frac{1}{2}K  +iA^5_j\gamma^j \gamma_5 \gamma^n)\Pi_+\, .
\lb{aaa}
\ee
One  finds  that
\be
E=-\frac{R}{4}+\frac{ie}{2}F_{ij}\hat{\gamma}^i\hat{\gamma}^j+i \gamma^5\nabla_i A^i_5 -A^5_i A^{5 i}\, .
\lb{a2}
\ee
Here $R$ is the scalar  curvature of $M_3$.
One defines $F_{ij}=\partial_i A_j-\partial_j A_i$.
The heat kernel coefficient that is needed for computation of the chiral anomaly is \cite{Vassilevich:2003xt}, \cite{GK}
\be
&&a_3(Q, \hat{D}_5^2)=\frac{1}{384(4\pi)}\int_{\partial M_3} \tr\{Q(x) (96\chi E+16\chi R 
-8\chi R_{nn}\nonumber \\
&& +(13\Pi_+-7\Pi_-)K^2+(2\Pi_+ +10\Pi_-)K_{ab}K^{ab} +96SK +192 S^2-12\bar{\nabla}_a\chi\bar{\nabla}^a\chi)\}\, .
\lb{a5}
\ee
Here $K_{ab}$ is the extrinsic curvature of the boundary,  $K$ is its trace and $\bar{\nabla}_a$ is covariant derivative along the boundary,
$\bar{\nabla}_a\chi=i\gamma_3\gamma^b K_{ab}+\omega_a\chi-\chi\omega_a$, see \cite{GK}.

In the case of chiral anomaly $Q(x)=\gamma_5\alpha(x)$. Then almost all traces either vanish or mutually cancel, leaving the only non-vanishing term as:
\be
\Tr (\gamma^5 \chi\hat{\gamma}^i\hat{\gamma}^j)F_{ij}=4F_{ij} \epsilon^{nij3}=4F_{ab} \epsilon^{ab}\, .
\lb{a6}
\ee
We use the relation:
\be
\Tr(\gamma^5\gamma^3 \gamma^n\hat{\gamma}^i\hat{\gamma}^j)=-4i \epsilon^{nij3}\, , 
\lb{a7}
\ee
where $n$ denotes the direction that is normal to boundary $\partial M_3$.  We use convention that in Minkowski spacetime with signature $(-,+,+,+)$ one has that $\epsilon^{0123}=1$.
The boundary epsilon symbol is defined as $\epsilon^{ab}=\epsilon^{nab3}$.
Putting things together we find that:
\be
a_3(\alpha(x)\gamma^5, \hat{D}^2)=\frac{ieN}{8\pi}\int_{\partial M_3}\alpha(x) \epsilon^{ab}F_{ab}\, .
\lb{a8}
\ee
This completes the computation of the anomaly.

\newpage


\end{document}